# Intuitionistic Fuzzy Linear Transformations

***Rajkumar Pradhan and Madhumangal Pal***

Department of Applied Mathematics with Oceanology and Computer Programming,
Vidyasagar University, Midnapore -- 721 102, India.
e-mail: mmpalvu@gmail.com

**Abstract.** In this paper, we discussed about the intuitionistic fuzzy linear transformations (IFLT) and shown that the set of all linear transformations $L(V)$ defined over an intuitionistic fuzzy vector space $V$ does not form an vector space. Here we determine the unique intuitionistic fuzzy matrix associated with an intuitionistic fuzzy linear transformation with respect to an ordered standard basis for an intuitionistic fuzzy vector space. We introduced the concept of the inverse of an IFLT.

## 1. Introduction

Uncertain or imprecise data are inherent and pervasive in many important applications in the areas such as economics, engineering, environment, social science, medical science and business management. There have been a great amount of research and applications in the literature concerning some special tools like probability theory, intuitionistic fuzzy set theory, rough set theory, vague set theory and interval mathematics to modeling uncertain data. However, all of these have theirs advantages as well as inherent limitations in dealing with uncertainties.

Initially, fuzzy set theory was proposed by Zadeh [24] as a means of representing mathematically any imprecise or vague system of information in the real world. In fuzzy set theory, there were no scope to think about the hesitation in the membership degrees which is arise in various real life situations. This situation is overcome in 1983 by invention of intuitionistic fuzzy sets [1]. Here it is possible to model hesitation and uncertainty by using an additional degree. The fuzzy matrices introduced first time by Thomson [21]. Kim and Roush [9] studied the canonical form of an idempotent fuzzy matrix. Hashimoto [4] studied the canonical form of a transitive fuzzy matrix. Xin [23] studied controllable fuzzy matrices. Hemashina et al. [5] investigated iterates of fuzzy circulant matrices. First time Pal [16] introduced intuitionistic fuzzy determinant. Im and Lee [6] studied on the determinant of square intuitionistic fuzzy matrix. Using the idea of intuitionistic fuzzy sets and intuitionistic fuzzy determinant Pal, Khan and Shaymal [15] introduced the intuitionistic fuzzy matrix and studied several properties on it. Pal and Shaymal [11] introduced the interval-valued fuzzy matrix and shown a lot of properties on it. Bhowmik and Pal [2] introduced some results on intuitionistic fuzzy matrix, intuitionistic circulant fuzzy matrix and generalized intuitionistic fuzzy matrix. Khan and Pal [19] introduced the concept of generalized inverse for intuitionistic fuzzy matrices, minus ordering and studied several properties of it. The notion of fuzzy relational equations based upon the max-min composition was first investigated by Sanchez [18]. In fuzzy algebra, fuzzy subspaces are basic concepts. They had been introduced by Katsaras and Liu [7] in 1977 as a generalization of the usual notion of vector spaces.

Meenakshi and Gandhimathi [13] introduce the concept of the linear transformation (LT) on intuitionistic fuzzy vector space (IFVS) and studied the several properties of it. In that paper they consider an intuitionistic fuzzy matrix (IFM) A as a cartesian product of two fuzzy matrices $A_\mu$ and $A_\nu$, where the fuzzy matrix $A_\mu$ represents the membership values and the fuzzy matrix $A_\nu$ represents the non-membership values of the elements of IFM A. They proved that two finite dimentional vector spaces over the fuzzy algebra are isomorphic if and only if their dimension are equal or if their exists a one-to-one and onto LT. They also proved that for an IFVS(V) over IF, L(V) is an algebra under multiplication defined by, $T_1T_2(x) = T_1(T_2(x))$ for all $T_1$, $T_2$ □ L(V). They also shown that for any T □ L(W) and the identity transformation I□L(W) the corresponding matrices [T] and [I] satisfy [T].[I]=[I].[T]=[T] under the max-min, min-max composition of intuitionistic fuzzy matrices. In this paper we study IFLT and its several properties using the properties of IFM directly. Meenakshi and gandhimathi [13] have claim that the set of LTs on IFVS V forms a vector space under addition and multiplication of LTs. But we have shown that, L(V) does not form an IFVS. Here we discussed about the the solution of the intuitionistic fuzzy relational equations which is not available in [13].

## 2. Preliminary Definitions

An intuitionistic fuzzy set forms an intuitionistic fuzzy algebra with two binary operations addition and multiplication if it satisfy the following properties:

(P1) Idempotence a+a=a, a.a=a.
(P2) Commutativity a+b=b+a, a.b=b.a.
(P3) Associativity a+(b+c)=(a+b)+c, a.(b.c)=(a.b).c.
(P4) Absorption a+(a.b)=a, a.(a+b)=a.
(P5) Distributivity a.(b+c)=(a.b)+(a.c), a+(b.c)=(a+b).(a+c).
(P6) Universal bounds a+0=a, a+1=1; a.0=0, a.1=a.

It can be shown that an intuitionistic fuzzy set is an intuitionistic fuzzy algebra with respect to max-min composition.

By an intuitionistic fuzzy matrix, we mean a matrix over intuitionistic fuzzy algebra, which is defined as follows.

**Definition 1. (Intuitionistic fuzzy matrices)[15]**

An intuitionistic fuzzy matrix (IFM) $A$ of order $m \times n$ is defined as $A=[x_{ij},\langle a_{ij\mu},a_{ij\nu}\rangle]_{m\times n}$ where $a_{ij\mu}$ , $a_{ij\nu}$ are called membership and non-membership values of $x_{ij}$ in $A$ , which maintains the condition $0\le a_{ij\mu}+a_{ij\nu}\le 1$ . For simplicity, we write $A=[x_{ij},a_{ij}]_{m\times n}$ or simply $[a_{ij}]_{m\times n}$ where $a_{ij}=\langle a_{ij\mu},a_{ij\nu}\rangle$ .

In arithmetic operations, only the values of $a_{ij\mu}$ and $a_{ij\nu}$ are needed so from here we only consider the values of $a_{ij}=\langle a_{ij\mu},a_{ij\nu}\rangle$ . All elements of an IFM are the members of $\langle F\rangle=\{\langle x,y\rangle : x,y\in[0,1]$ and $0\le x+y\le 1\}$ .

**Definition 2. (Fuzzy vector space)[7]**

A fuzzy vector space $(V,\phi)$ is a classical vector space $V$ with a map $\phi:V\to[0,1]$ satisfying the following conditions.

$(i)$ $\phi(a+b)\ge\min\{\phi(a),\phi(b)\}$.

$(ii)$ $\phi(-a)=\phi(a)$.

$(iii)$ $\phi(0)=1$.

$(iv)$ $\phi(\alpha a)\ge\phi(a)$ for all $a,b\in V$ and $\alpha\in F$.

The above conditions can be combined to a single condition as $\phi(\alpha a+\beta b)\ge\min\{\phi(a),\phi(b)\}$ for all $a,b\in V$ and $\alpha,\beta\in F$ .

The definition of Fuzzy vector space in the above form motivates us to defined the Intuitionistic fuzzy vector space as follows:

**Definition 3. (Intuitionistic fuzzy vector space)**

Let $A=(\mu_A,\nu_A)$ be an intuitionistic fuzzy set of a classical vector space $V$ over $F$ . For any $x,y\in V$ and $\alpha,\beta\in F$ , if it satisfy

$\mu_A(\alpha x+\beta y)\ge\min\{\mu_A(x),\mu_A(y)\}$ and $\nu_A(\alpha x+\beta y)\le\max\{\nu_A(x),\nu_A(y)\}$

then $A$ is called an intuitionistic fuzzy subspace of $V$ .

Let $V_n$ denotes the set of all $n$ -tuples $(\langle x_{1\mu},x_{1\nu}\rangle,\langle x_{2\mu},x_{2\nu}\rangle,\ldots,\langle x_{n\mu},x_{n\nu}\rangle)$ over $F$ . An element of $V_n$ is called an intuitionistic fuzzy vector (IFV) of dimension $n$, where $x_{i\mu}$ and $x_{i\nu}$ are the membership and non-membership values of the component $x_i$ .

**Definition 4. (Dependence of IFVs)**

A set $S$ of intuitionistic fuzzy vectors is independent if and only if each element of $S$ can not be expressed as a linear combination of other elements of $S$ , that is, no element $s\in S$ is a linear combination of $S\setminus\{s\}$ .

A vector $\alpha$ may be expressed by some other vectors. If it is possible then the vector $\alpha$ is called dependent otherwise it is called independent. These terminologies are similar to classical vectors.

The examples of independent and dependent set of vectors are given below.

**Example 1.** *Let* $S=\{a_1,a_2,a_3\}$ *be a subset of* $V_3$*, where*

$a_1 = (\langle 0.8,0.2\rangle, \langle 0.6,0.3\rangle, \langle 0.4,0.3\rangle), a_2 = (\langle 0.5,0.3\rangle, \langle 0.5,0.1\rangle, \langle 0.4,0.2\rangle)$ *and*
$a_3 = (\langle 0.7,0.3\rangle, \langle 0.7,0.2\rangle, \langle 0.9,0.1\rangle)$.

Here the set $S$ is an independent set. If not then $a_1 = \alpha a_2 + \beta a_3$ for $\alpha, \beta \in F$.

So, $a_1 = \alpha(\langle 0.5,0.3\rangle, \langle 0.5,0.1\rangle, \langle 0.4,0.2\rangle) + \beta(\langle 0.7,0.3\rangle, \langle 0.7,0.2\rangle, \langle 0.9,0.1\rangle)$.

It is not possible to find any $\alpha, \beta \in F$ such that the corresponding coefficients on both sides will be equal. That is, $a_1 \neq \alpha a_2 + \beta a_3$. Similarly, $a_2 \neq \alpha a_1 + \beta a_3$ and $a_3 \neq \alpha a_2 + \beta a_1$. So the set $S$ is independent.

Let $S = \{a_1, a_2\}$ be a subset of $V_3$, where $a_1 = (\langle 0.7,0.3\rangle, \langle 0.5,0.3\rangle, \langle 0.6,0.3\rangle)$ and $a_2 = (\langle 0.8,0.2\rangle, \langle 0.5,0.1\rangle, \langle 0.6,0.2\rangle)$ Here $a_1 = ca_2$ for $c = 0.7$. So $S$ is a dependent set.

**Definition 5.** **(Basis)** Let $W$ be an intuitionistic fuzzy subspace of $V_n$ and $S$ be a subset of $W$ such that the elements of $S$ are independent. If every element of $W$ can be expressed uniquely as a linear combination of the elements of $S$, then $S$ is called a basis of intuitionistic fuzzy subspace $W$.

## 3. Intuitionistic Fuzzy Relational Equation

In this section, we describe how intuitionistic fuzzy relational equation is associated with the concept of composition of binary relations. Any relation between two sets $X$ and $Y$ is known as a binary relation. For any $x \in X$ and $y \in Y$, by $xRy$ we mean that, $x$ is related to $y$ under the given relation $R$. For intuitionistic fuzzy relation $R(x, y)$ we write the membership and non-membership value of $x$ in relation with $y$ under $R$ denoted by $\mu_R(x, y) = \alpha_\mu$ and $\nu_R(x, y) = \alpha_\nu$ and represented as $\langle \alpha_\mu, \alpha_\nu \rangle$. A binary intuitionistic fuzzy relation $R$ can be represented as an intuitionistic fuzzy membership matrix $M_R$.

**Example 2.** *Let $R$ be an intuitionistic fuzzy relation between two sets $X = \{Amal, Kamal, Bimal\}$ and $Y = \{Mr. Ram, Mr. Shyam\}$ that represents the relational concept ``that the elements of set $X$ cast their vote for a particular candidate or not of the set $Y$ ". This relation can be represented by the following intuitionistic fuzzy matrix,*

$$M_R = \begin{array}{c} \\ \text{Mr. Ram} \\ \text{Mr. Shaym} \end{array} \begin{array}{c} \begin{array}{ccc} \text{Amal} & \text{Kamal} & \text{Bimal} \end{array} \\ \begin{bmatrix} \langle 0.4,0.3\rangle & \langle 0.5,0.5\rangle & \langle 0.6,0.2\rangle \\ \langle 0.3,0.5\rangle & \langle 0.4,0.3\rangle & \langle 0.3,0.5\rangle \end{bmatrix} \end{array}$$

The composition of two intuitionistic fuzzy binary relations $A(X, Y)$ and $B(Y, Z)$ are defined in terms of the max-min operation denoted as $A \circ B$ on the corresponding intuitionistic fuzzy matrices of $A$ and $B$. Let us consider the intuitionistic fuzzy binary relations $A(X, Y)$, $B(X, Y)$ and $R(X, Z)$ defined on the sets $X = \{x_i / i \in N_s\}$, $Y = \{y_j / j \in N_m\}$ and $Z = \{z_k / k \in N_m\}$, where $N_m$ is the set of all positive integers 1 to $m$. Let the corresponding intuitionictic fuzzy matrices be denoted by $A=(a_{ij})=\langle a_{ij\mu},a_{ij\nu}\rangle$ , $B=(b_{jk})=\langle b_{jk\mu},b_{jk\nu}\rangle$ and $R=(r_{ik})=\langle r_{ik\mu},r_{ik\nu}\rangle$ respectively. The composition $R(X,Z)$ of $A(X,Y)$ and $B(Y,Z)$ is given by intuitionistic fuzzy matrix equation that is,

$$AB = R. \tag{1}$$

or, $$(\langle \max_j \{\min(a_{ij\mu},b_{jk\mu})\}, \min_j \{\max(a_{ij\nu},b_{jk\nu})\}\rangle) = \langle r_{ik\mu},r_{ik\nu}\rangle \tag{2}$$

where $i\in N_s$, $j\in N_m$ and $k\in N_n$.

The intuitionistic fuzzy matrix equation (1) represents $s\times n$ simultaneous equations of the form of the equation (2). When two of the components in each of the equations are given and one is unknown, these equations are called the *intuitionistic fuzzy relational equations*. In equation (1) when intuitionistic fuzzy matrices $A$ and $B$ are known then $R$ can be determined uniquely be performing the max-min operations on $A$ and $B$.

Let us assume that a pair of IFMs $R$ and $B$ are given, we wish to determine the set of all particular matrices $A$ that satisfy the the equation (1). Each particular IFM $A$ that satisfies the equation (1) is called its solution and the set $\Omega(B,R)=\{A/AB=R\}$ denotes the solution set.

In the further discussion in this section we consider the intuitionistic fuzzy matrix equations of the form $Ax=b$ (3)
with $x=[\langle x_{j\mu},x_{j\nu}\rangle/j\in N_m]^T$ , $b=[\langle b_{k\mu},x_{k\nu}\rangle/k\in N_n]^T$ and $A\in IF_{m\times n}$ and the solution set $\Omega(A,b)$.

**Definition 6.** *Any element $\hat{x}$ of $\Omega(A,b)$ is called a solution of the equation $Ax=b$ if $\hat{x}=[\langle x_{j\mu},x_{j\nu}\rangle/j\in N_m]^T$ be defined as $\hat{x}=\min\sigma(a_{jk},b_k)$ where,*

$$\sigma(a_{jk},b_k)=\begin{cases} b_k & \text{if } a_{jk}>b_k;\\ \langle 1,0\rangle & \text{otherwise.}\end{cases}$$

In case of intuitionistic fuzzy relational equation of the form $Ax=b$ for any arbitrary $A$ and $b$ there may or may not exists a solution $\hat{x}$.

If $A$ is regular that is, $A$ has a g-inverse [17], then there exists a solution of the equation $Ax=b$.

**Example 3.** *Given* $A=\begin{bmatrix}\langle 0.8,0.2\rangle & \langle 0.6,0.3\rangle\\ \langle 0.6,0.4\rangle & \langle 0.7,0.2\rangle\end{bmatrix}$ *and b=* $[\langle 0.7,0.3\rangle \quad \langle 0.6,0.3\rangle]^T$.
*Find out one solution of the equation $Ax=b$.*

**Solution:** From Definition (6),

$$x_1=\min\sigma(a_{1k},b_k)=\min\{\sigma(a_{11},b_1),\sigma(a_{12},b_2)\}=\min\{\langle 0.7,0.3\rangle,\langle 1,0\rangle\}=\langle 0.7,0.3\rangle$$

$$x_2=\min\sigma(a_{2k},b_k)=\min\{\sigma(a_{21},b_1),\sigma(a_{22},b_2)\}=\min\{\langle 1,0\rangle,\langle 0.6,0.3\rangle\}=\langle 0.6,0.3\rangle.$$

Hence one of the solution for the equation $Ax=b$ is $\hat{x}=[\langle 0.7,0.3\rangle,\langle 0.6,0.3\rangle]$.

## 4. Intuitionistic Fuzzy Linear Transformation (IFLT)

In this section, we study the concept of linear transformation on intuitionistic fuzzy vector space and show that the set of all linear transformation between two intuitionistic fuzzy vector spaces does not form a vector space under the fuzzy operations. IFLT is very useful to describe the projection of one vector into another vector and to rotate a vector in some intuitionistic fuzzy vector space.

**Definition 7.** *Let $V$ and $W$ be two vector spaces over the intuitionistic fuzzy algebra IF. A mapping $T$ of $V$ into $W$ is called a linear transformation if for any $x, y \in V$ and $\alpha$ □ IF,*

(i). $T(x+y) = T(x) + T(y)$ and

(ii). $T(\alpha x) = \alpha . T(x)$.

This linear transformation is called IFLT.

**Example 4.** *Let $V_3$ and $V_2$ be vector spaces over IF. The mapping $T: V_3 \to V_2$ defined as $T(x_1, x_2, x_3) = (x_1, x_2)$ is a linear transformation.*

**Theorem 1.** *Let $V$ be a vector space over IF and $L(V)$ be the set of all linear transformations defined on $V$. Then $L(V)$ is closed under addition and multiplication defined by*

$(T_1 + T_2)(x) = T_1(x) + T_2(x)$

$(\alpha T_1).x = \alpha . T_1(x)$ for all $T_1, T_2 \in V, x \in V$ and α □ *IF.*

**Proof:** For $x, y \in V$,

$$
\begin{aligned}
(T_1+T_2)(x+y) &= T_1(x+y) + T_2(x+y) \\
&= T_1(x) + T_1(y) + T_2(x) + T_2(y) \\
&= (T_1+T_2)(x) + (T_1+T_2)(y), \text{ for all } T_1, T_2 \text{ in } L(V).
\end{aligned}
$$

Again,

$$
\begin{aligned}
(T_1+T_2)(\alpha x) &= T_1(\alpha x) + T_2(\alpha x) = \alpha T_1(x) + \alpha T_2(x) = \alpha(T_1(x) + T_2(x)) \\
&= \alpha.(T_1+T_2)(x), \textit{ for all } T_1, T_2 \in L(V) \textit{ and } \alpha \in \text{IF}.
\end{aligned}
$$

Thus, $T_1 + T_2 \in L(V)$ for all $T_1, T_2 \in L(V)$.

For, α □ *IF* and $T \in L(V)$,

$(\alpha T)(x+y) = \alpha(T(x+y)) = \alpha(T(x) + T(y)) = \alpha . T(x) + \alpha . T(y)$.

and $(\alpha T)(\beta x) = \alpha(T(\beta x)) = \alpha(\beta T(x)) = \beta((\alpha T)(x))$.

Hence, $\alpha T \in L(V)$.

So $L(V)$ is closed under addition and multiplication.

**Theorem 2.** *For $T_1, T_2, T_3 \in L(V)$ and α,β □ IF the following properties are hold.*

(i) $T_1 + T_2 = T_2 + T_1$.

(ii) $(T_1 + T_2) + T_3 = T_1 + (T_2 + T_3)$.

(iii) $(\alpha\beta)T_1 = \alpha(\beta T_1)$.

(iv) $(\alpha + \beta).T_1 = \alpha . T_1 + \beta . T_1$.

(v) $\alpha(T_1+T_2)=\alpha.T_1+\alpha.T_2$.

(vi) If $T(x)=x$, then $T.T_1=T_1$ for all $T_1\in L(V)$.

(vii) If $T(x)=0$, then $T.T_1=0$ for all $T_1\in L(V)$.

**Proof:** (i) For any $x\in V$,

$(T_1+T_2)(x)=T_1(x)+T_2(x)=T_2(x)+T_1(x)=(T_1+T_2)(x)$.

Hence, $T_1+T_2=T_2+T_1$.

(ii) For any $x\in V$,

$$\begin{aligned}((T_1+T_2)+T_3)(x) &= (T_1+T_2)x+T_3(x)=T_1(x)+T_2(x)+T_3(x)=T_1(x)+(T_2(x)+T_3(x))\\ &= (T_1+(T_2+T_3))(x).\end{aligned}$$

Hence, $(T_1+T_2)+T_3=T_1+(T_2+T_3)$.

(iii) For any $x\in V$,

$$((\alpha\beta)T_1)(x)=\alpha\beta(T_1(x))=\alpha(\beta T_1(x))=\alpha((\beta T_1)(x))=(\alpha(\beta T_1))(x).$$

Hence, $(\alpha\beta)T_1=\alpha(\beta T_1)$.

(iv) For, α,β  *IF*

$$\begin{aligned}((\alpha+\beta)T_1)(x) &= (\alpha+\beta)(T_1(x))=\alpha(T_1(x))+\beta(T_1(x))=\alpha.T_1(x)+\beta.T_1(x)\\ &= (\alpha T_1+\beta T_1)(x),\forall x\in V.\end{aligned}$$

Hence, $(\alpha+\beta)T_1=\alpha.T_1+\beta.T_1$.

(v) For, α  *IF*

$$\begin{aligned}\alpha(T_1+T_2)(x) &= \alpha((T_1+T_2)(x))=\alpha(T_1(x)+T_2(x))=\alpha.T_1(x)+\alpha.T_2(x)\\ &= (\alpha T_1+\alpha T_2)(x),\forall x\in V.\end{aligned}$$

Hence, $\alpha(T_1+T_2)=\alpha.T_1+\alpha.T_2$.

(vi) For the linear transformation $T(x)=x$, $T.T_1=T_1$, for all $T_1\in L(V)$.

(vii) For the linear transformation $T(x)=0$, $T.T_1=0$, for all $T_1\in L(V)$.

**Example 5.** *Let* $T_1(x)=\langle 0.6,0.4\rangle(x)$, $T_2(x)=\langle 0.5,0.5\rangle(x)\in L(V_2)$ *and* $x=(\langle 0.7,0.2\rangle,\langle 0.6,0.3\rangle)\in V_2$.

Then it can be verified that $T_1+T_2=T_2+T_1$.

For $T_3(x)=x\in L(V_2)$, it can be shown that $(T_1+T_2)+T_3=T_1+(T_2+T_3)$.

Let $\alpha=\langle 0.6,0.2\rangle$ and $\beta=\langle 0.5,0.3\rangle$ be two elements of *IF*.

Then, it can also be shown that $(\alpha\beta)T_1=\alpha(\beta T_1)$, $(\alpha+\beta)T_1=\alpha.T_1+\beta.T_1$ and $\alpha(T_1+T_2)=\alpha.T_1+\alpha.T_2$.

For $T(x)=x\in L(V_2)$ and $T(x)=0=(\langle 0,0\rangle,\langle 0,0\rangle)\in L(V_2)$,

it can also be verified that, $T.T_1=T_1$ and $T.T_1=T$.

*By the following example we shown that* $L(V_n)$ *does not form an vector space*

*Let* $T(x)=(\langle 0,0\rangle,\langle 0,0\rangle,\ldots,\langle 0,0\rangle)$ *and* $T(x)=x\in L(V_n)$ *for any* $x\in V_n$. *Then,*

$$
\begin{aligned}
(T+T_1)(x) &= T(x)+T_1(x) \\
&= (\langle 0,0\rangle,\langle 0,0\rangle,\ldots,\langle 0,0\rangle)+(\langle x_{1\mu},x_{1\nu}\rangle,\langle x_{2\mu},x_{2\nu}\rangle,\ldots,\langle x_{n\mu},x_{n\nu}\rangle) \\
&= (\langle x_{1\mu},0\rangle,\langle x_{2\mu},0\rangle,\ldots,\langle x_{2\mu},0\rangle) \\
&\neq T_1(x)
\end{aligned}
$$

So $L(V_n)$ does not form a vector space over the intuitionistic fuzzy algebra IF under the addition and fuzzy multiplication defined in Theorem 1.

## 5. Intuitionistic Fuzzy Linear Combinations

We know that any vector in the subspace over IF can be expressed uniquely as a intuitionistic fuzzy linear combination of its standard basis vectors. In this section, we find the intuitionistic fuzzy linear combination of the basis vectors.

**Definition 8. (Standard basis)**

A basis $B$ of an intuitionistic fuzzy vector space $W$ is a standard basis if and only if whenever $b_i=\sum_{j=1}^{n}a_{ij}b_j$ for $b_i,b_j\in B$ and $a_{ij}\in[0,1]$ then $a_{ii}b_i=b_i$.

**Theorem 3.** *Let $\{c_1,c_2,\ldots,c_n\}$ be the standard basis of the subspace $W$ of $V_n$. In the intuitionistic fuzzy linear combination of the basis vector $c_i$, the $i$ th coefficient if $c_i$ is $\langle 1,0\rangle$.*

**Proof:** Since $c_i\in V_n$, let $c_i=(\langle c_{i1\mu},c_{i1\nu}\rangle,\langle c_{i2\mu},c_{i2\nu}\rangle,\ldots,\langle c_{in\mu},c_{in\nu}\rangle)$ for each $i=1$ to $n$. Then the intuitionistic fuzzy linear combination of the basis vector $c_i$ in terms of the standard basis $\{c_1,c_2,\ldots,c_n\}$ can be written as

$$
\begin{aligned}
& c_i = c_1x_1+c_2x_2+\ldots+c_nx_n \\
& or,(\langle c_{i1\mu},c_{i1\nu}\rangle,\langle c_{i2\mu},c_{i2\nu}\rangle,\ldots,\langle c_{in\mu},c_{in\nu}\rangle) \\
& =(\langle c_{11\mu},c_{11\nu}\rangle,\langle c_{12\mu},c_{12\nu}\rangle,\ldots,\langle c_{1n\mu},c_{1n\nu}\rangle)\langle x_{1\mu},x_{1\nu}\rangle \\
& +(\langle c_{21\mu},c_{21\nu}\rangle,\langle c_{22\mu},c_{22\nu}\rangle,\ldots,\langle c_{2n\mu},c_{2n\nu}\rangle)\langle x_{2\mu},x_{2\nu}\rangle \\
& +\ldots\ldots\ldots\ldots\ldots\ldots\ldots\ldots\ldots\ldots\ldots\ldots\ldots\ldots\ldots\ldots \\
& +(\langle c_{i1\mu},c_{i1\nu}\rangle,\langle c_{i2\mu},c_{i2\nu}\rangle,\ldots,\langle c_{in\mu},c_{in\nu}\rangle)\langle x_{i\mu},x_{i\nu}\rangle \\
& +\ldots\ldots\ldots\ldots\ldots\ldots\ldots\ldots\ldots\ldots\ldots\ldots\ldots\ldots\ldots\ldots \\
& +(\langle c_{n1\mu},c_{n1\nu}\rangle,\langle c_{n2\mu},c_{n2\nu}\rangle,\ldots,\langle c_{nn\mu},c_{nn\nu}\rangle)\langle x_{n\mu},x_{n\nu}\rangle.
\end{aligned}
$$

This can be expressed as $Ax=b$ where,

$b=(\langle c_{i1\mu},c_{i1\nu}\rangle,\langle c_{i2\mu},c_{i2\nu}\rangle,\ldots,\langle c_{in\mu},c_{in\nu}\rangle)^T$ and

$x=(\langle x_{1\mu},x_{1\nu}\rangle,\langle x_{2\mu},x_{2\nu}\rangle,\ldots,\langle x_{n\mu},x_{n\nu}\rangle)^T$ and

$$A=\begin{bmatrix} \langle c_{11\mu},c_{11\nu}\rangle & \langle c_{12\mu},c_{12\nu}\rangle & \dots & \langle c_{1n\mu},c_{1n\nu}\rangle \\ \langle c_{21\mu},c_{21\nu}\rangle & \langle c_{22\mu},c_{22\nu}\rangle & \dots & \langle c_{2n\mu},c_{2n\nu}\rangle \\ \dots\dots\dots & \dots\dots\dots & \dots & \dots\dots\dots \\ \langle c_{i1\mu},c_{i1\nu}\rangle & \langle c_{i2\mu},c_{i2\nu}\rangle & \dots & \langle c_{in\mu},c_{in\nu}\rangle \\ \dots\dots\dots & \dots\dots\dots & \dots & \dots\dots\dots \\ \langle c_{n1\mu},c_{n1\nu}\rangle & \langle c_{n2\mu},c_{n2\nu}\rangle & \dots & \langle c_{nn\mu},c_{nn\nu}\rangle \end{bmatrix}.$$

Let us find the solution of the intuitionistic fuzzy relational equation $Ax=b$ by using Definition 6. So the $i$ th component of $x_i$ of the solution $x=(x_1,x_2,\dots,x_n)$ is defined by

$$\begin{aligned} x_i &= \min\{\sigma(a_{i1},b_1),\sigma(a_{i2},b_2),\dots,\sigma(a_{in},b_n)\} \\ &= \min\{\sigma(c_{i1},b_1),\sigma(c_{i2},b_2),\dots,\sigma(c_{in},b_n)\} \\ &= \min\{\sigma(\langle c_{i1\mu},c_{i1\nu}\rangle,\langle c_{i1\mu},c_{i1\nu}\rangle),\sigma(\langle c_{i2\mu},c_{i2\nu}\rangle,\langle c_{i2\mu},c_{i2\nu}\rangle),\dots,\sigma(\langle c_{in\mu},c_{in\nu}\rangle,\langle c_{in\mu},c_{in\nu}\rangle)\} \\ &= \min\{\langle 1,0\rangle,\langle 1,0\rangle,\dots,\langle 1,0\rangle\} \\ &= <1,0> \end{aligned}$$

## 6. Intuitionistic Fuzzy Matrices Associated With IFLT

Let $B=\{c_1,c_2,\dots,c_n\}$ be an order standard basis for a subspace $W$ of intuitionistic fuzzy vector space $V_n$. Then each vector in $W$ is uniquely expressible as an intuitionistic fuzzy linear combination of $c_i$ s.

Let $T$ be a linear transformation on $W$ to itself, then for each $c_j=\langle c_{j\mu},c_{j\nu}\rangle$ the vector $T(c_j)=T(\langle c_{j\mu},c_{j\nu}\rangle)$ is in $W$. Then $T(c_j)=T(\langle c_{j\mu},c_{j\nu}\rangle)=\sum_{i=1}^{n}\langle \alpha_{ij\mu},\alpha_{ij\nu}\rangle\langle c_{i\mu},c_{i\nu}\rangle$ is the intuitionistic fuzzy linear combination interms of the standard basis vectors. Now we construct the matrix $[T]$, whose $j$ th column is the transpose of the intuitionistic row vector $(\langle a_{1j\mu},a_{1j\nu}\rangle,\langle a_{2j\mu},a_{2j\nu}\rangle,\dots,\langle a_{nj\mu},a_{nj\nu}\rangle)$. Since $a_{ij}=\langle a_{ij\mu},a_{ij\nu}\rangle$ are uniquely determined by $T(c_j)=T(\langle c_{j\mu},c_{j\nu}\rangle)$ for each $j$, $[T]$ is the unique intuitionistic fuzzy matrix corresponding to the linear transformation $T$ on $L(W)$ with respect to the unique ordered standard basis $B$ of the subspace $W$. Let us denote it as $[T]_B$

**Example 6.** *Let* $S=\{\langle x_\mu,x_\nu\rangle,\langle y_\mu,y_\nu\rangle,\langle z_\mu,z_\nu\rangle \mid x_\nu=y_\nu=z_\nu\}$ *be a subspace of the intuitionistic fuzzy vector space* $V_3$*. Now,* $B_1=\{e_1,e_2,e_3\}=\{(\langle 1,0\rangle,\langle 0,0\rangle,\langle 0,0\rangle),(\langle 0,0\rangle,\langle 1,0\rangle,\langle 0,0\rangle),(\langle 0,0\rangle,\langle 0,0\rangle,\langle 1,0\rangle)\}$ *is ordered standard basis for the subspace* $S$.

Similarly, $W=\{\langle x_\mu,x_\nu\rangle,\langle y_\mu,y_\nu\rangle \mid x_\nu=y_\nu\}$ be a subspace of the fuzzy vector space $V_2$. Now, $B_2 = \{e_{1'}, e_{2'}\} = \{(\langle 1,0\rangle, \langle 0,0\rangle), (\langle 0,0\rangle, \langle 1,0\rangle)$ is a ordered standard basis for the subspace $W$.

Let $T: S \to W$ such that $T(x_1, x_2, x_3) = (x_1, x_2)$ for every $x \in S$ be the linear transformation from $S$ to $W$.Then,

$$\begin{aligned} T(e_1) &= (\langle 1,0\rangle, \langle 0,0\rangle) \\ &= \langle 1,0\rangle(\langle 1,0\rangle, \langle 0,0\rangle) + \langle 0,0\rangle(\langle 0,0\rangle, \langle 1,0\rangle) \\ T(e_2) &= (\langle 0,0\rangle, \langle 1,0\rangle) \\ &= \langle 0,0\rangle(\langle 1,0\rangle, \langle 0,0\rangle) + \langle 1,0\rangle(\langle 0,0\rangle, \langle 1,0\rangle) \\ T(e_3) &= (\langle 0,0\rangle, \langle 0,0\rangle) \\ &= \langle 0,1\rangle(\langle 1,0\rangle, \langle 0,0\rangle) + \langle 0,0\rangle(\langle 0,0\rangle, \langle 1,0\rangle). \end{aligned}$$

Hence the intuitionistic fuzzy matrix for the intuitionistic fuzzy linear transformation with respect to the above ordered standard basis is,

$$[T] = \begin{bmatrix} \langle 1,0\rangle & \langle 0,0\rangle & \langle 0,1\rangle \\ \langle 0,0\rangle & \langle 1,0\rangle & \langle 0,0\rangle \end{bmatrix}$$

**Definition 9.** *Let $T$ be a linear transformation defined on the intuitionistic fuzzy vector space $V$. Then for $Y \in L(V)$ is said to be the inverse linear transformation of $T$ if $[T][Y][T] = [T]$, where $[T]$ and $[Y]$ are the matrices associated with the corresponding IFLTs.*

**Example 7.** *Let $V$ be a subspace generated by the standard basis $B = \{c_1, c_2\}$, where $c_1 = (\langle 0.5,0\rangle, \langle 0.5,0\rangle)$ and $c_2 = (\langle 0,1\rangle, \langle 1,0\rangle)$. Let $T: V \to V$ such that $T(x) = \langle 0.5,0.3\rangle(x)$, for every $x \in V$ be the intuitionistic fuzzy linear transformation on V.*

So the associative matrix of the intuitionistic fuzzy linear transformation $T(x)$ with respect to the standard basis $B$ is $[T] = \begin{bmatrix} \langle 0.6,0.3\rangle & \langle 0,1\rangle \\ \langle 0.4,0.4\rangle & \langle 0.5,0.3\rangle \end{bmatrix}$

Now, for the associative matrix $[Y] = \begin{bmatrix} \langle 0.7,0.3\rangle & \langle 0,1\rangle \\ \langle 0.3,0.5\rangle & \langle 0.5,0.3\rangle \end{bmatrix}$ of the intuitionistic fuzzy linear transformation $Y(x)$ with respect to the same standard basis $B$, $[T][Y][T] = [T]$ holds.

In this case,

$$
\begin{aligned}
Y(c_1) &= \langle 0.7,0.3\rangle(\langle 0.5,0\rangle,\langle 0.5,0\rangle)+\langle 0.3,0.5\rangle(\langle 0,1\rangle,\langle 1,0\rangle)\\
&= (\langle 0.5,0.3\rangle,\langle 0.5,0.3\rangle)+(\langle 0,1\rangle,\langle 0.3,0.5\rangle)\\
&= (\langle 0.5,0.3\rangle,\langle 0.5,0.3\rangle)\\
&= \langle 0.5,0.3\rangle(\langle 0.5,0\rangle,\langle 0.5,0\rangle)\\
Y(c_2) &= \langle 0,1\rangle(\langle 0.5,0\rangle,\langle 0.5,0\rangle)+\langle 0.5,0.3\rangle(\langle 0,1\rangle,\langle 1,0\rangle)\\
&= (\langle 0,1\rangle,\langle 0.5,0.3\rangle)\\
&= \langle 0.5,0.3\rangle(\langle 0,1\rangle,\langle 1,0\rangle)
\end{aligned}
$$

Hence, $Y(x)=\langle 0.5,0.3\rangle(x)$ for all $x \in V$. That is, $T(x)$ is self inverse here.